\documentclass[12pt]{article}

\usepackage{graphicx}

\begin{document}

\begin{center}
{\bf Three-Dimensional Nonlinear Stokes$-$Mueller Polarimetry } \\
\vspace{5mm} 
Serguei Krouglov\footnote{E-mail: serguei.krouglov@utoronto.ca},
Virginijus Barzda\footnote{E-mail: virgis.barzda@utoronto.ca}
\underline{}
\vspace{3mm}

\textit{Department of Physics, University of Toronto, \\60 St. Georges St.,
Toronto, ON M5S 1A7, Canada\\
Department of Chemical and Physical Sciences,\\ University of Toronto Mississauga,\\
3359 Mississauga Rd. N., Mississauga, ON L5L 1C6, Canada}

\end{center}
\vspace{5mm}
Codes{(030.0030) Coherence and statistical optics, (260.2130) Stokes-Mueller ellipsometry and polarimetry, (260.5430) Polarization, (120.6710) Nonlinear Susceptibility, (270.1670) Coherent optical effects, (290.5855) Scattering, polarization, (190.0190) Nonlinear optics, (190.4160) Multiharmonic generation, (190.4380) Nonlinear optics, four-wave mixing, (180.4315) Nonlinear microscopy, (170.3880) Medical and biological imaging.}

\begin{abstract}
The formalism is developed for a three-dimensional ($3D$) nonlinear Stokes-Mueller polarimetry that describes a method of acquiring a complete complex valued 3D nonlinear susceptibility tensor of a material. The expressions are derived for generalized $3D$ linear and nonlinear Stokes vectors, and the corresponding nonlinear Mueller matrix.The coherency-like Hermitian square matrix $X$ of susceptibilities is introduced, which is derived from the nonlinear Mueller matrix. The $X$-matrix is characterized by the index of depolarization. Several decompositions of the $X$-matrix are introduced that provide a possibility to obtain nonlinear susceptibility tensors of constituting materials in the heterogeneous media. The $3D$ nonlinear Stokes-Mueller polarimetry formalism can be applied for three and higher wave mixing processes. The $3D$ polarimetric measurements can be used for structural investigations of materials, including heterogeneous biological structures. The $3D$ polarimetry is applicable for nonlinear microscopy with high numerical aperture objectives.
\end{abstract}

\section{Introduction}

The molecular organization and symmetry of materials can be probed by measuring polarization of nonlinear optical signals as a function of polarization states of incoming fundamental radiation \cite{Kruglov}. Polarimetric measurements of second harmonic generation (SHG), sum-frequency generation (SFG), third harmonic generation (THG), coherent anti-Stokes Raman scattering (CARS) and stimulated Raman scattering (SRS) have been applied for studying oriented molecules \cite{Verbiest}, \cite{Simpson}. The nonlinear polarimetry proved to be very beneficial in microscopy for investigating ultrastructural organization of materials and biological tissue beyond the diffraction limit \cite{Clays}
%, \cite{Maker}, \cite{Shi}, \cite{Zhuang} \cite{Tuer}, \cite{Loew}, \cite{Plotnikov}, \cite{Nucciotti}, \cite{Vanzi},
- \cite{Kontenis1}.
Stokes-Mueller and Jones formalisms can be used to characterize the nonlinear light-matter interactions \cite{Kliger}, \cite{Shurcliff}. The Jones approach employs expressions of complex electric fields to describe the pure polarized states and, therefore, it has limited applications for cases with heterogeneous scattering media. On the other hand, the Stokes vector is expressed via intensities and can describe partially polarized light interactions. Therefore, the Stokes-Mueller formalism is well suited for describing nonlinear light-matter interactions in heterogeneous scattering media such as biological tissue \cite{Lien} - \cite{Mazumder2}.
%, \cite{Mazumder}, \cite{Mazumder1},

The Stokes-Mueller polarimetry measurements are performed by considering the Stokes vectors that describe two dimensional ($2D$) polarization states. However, numerous polarimetry applications, including polarization microscopy, employ non-collinear beam geometries and  highly converging/diverging beams, as in the case of imaging with high numerical aperture (NA) objectives. Only recently a three-dimensional ($3D$) Stokes-Mueller formalism has been developed for the linear interaction regime.
The formalism is suited for investigations where there is no well-defined propagation direction of light \cite{Samson} - \cite{Brosseau_0}.

In this paper we develop $3D$ nonlinear Stokes-Mueller formalism to aid in the polarimetric nonlinear microscopy and non-collinear polarimetry measurements.
The $3D$ nonlinear Stokes-Mueller polarimetry formalism is developed with the SFG and SHG polarimetric measurements in mind. This formalism makes use of $3D$ Stokes vector having $9$ components and representing linearly interacting electric fields \cite{Brosseau} - \cite{Gil}. The approach allows us to consider the situation when the plane-wave approximation is not perfectly valid. This can be realised if the incident or detected light from a microscope are not measured in the far-field. The example is the use of high NA lenses. Another case for using new $3D$ polarimetry is when two incident beams are not parallel and propagate to the sample.

In Section 2, we introduce $3D$ linear and double Stokes vectors. The elements of $9\times 9$ coherency matrix are expressed through elements of $3D$ linear Stokes vector. The corresponding expression for double Mueller matrix, containing $9\times 81$ components, is introduced. The description is provided on how to calculate the Mueller matrix elements from the polarimetric measurements. In Section 3, the coherency-like matrix $X$ is introduced, which allows us to transform $9\times 81$ Mueller matrix into $27\times 27$
square matrix $X$ \cite{Kruglov}. The formalism of index of depolarization is introduced that characterizes the square matrix $X$. Different decompositions of $X$-matrix are considered in Section 4. In Appendix A, we describe the parametrization of the SHG outgoing radiation coherency matrix. We derive the expression for the double Stokes vector $S$ for the degenerate case of SHG via generalized $3D$ Stokes vector $s$ in Appendix B. In Appendix C, the
transformations of Mueller and $X$-matrices under rotation of Cartesian coordinate system are presented.
\begin{table}[ht]
\caption{The notations}
\centering
\begin{tabular}{ccccccccccccc }\\[1ex]
 \hline
 $s'$ & the linear $9\times 1$ Stokes vector\\
\hline
${\cal M}$ & the Mueller matrix  \\[1ex]
\hline
$S$ & the nonlinear Stokes vector\\
\hline
$\Omega'$ & $3$-component linear vector\\
\hline
$C$ & $3\times 3$ coherency matrix  \\
\hline
$\lambda_A$ & Gell-Mann matrices\\
\hline
${\cal P}_i$ & the polarization vector\\
\hline
$\Psi$ & $9$-component state vector\\
\hline
$\eta_N$ & the base matrices \\
\hline
$X$ &the correlation matrix\\
\hline
$P_D$ & the index of depolarization\\
\hline
$P_i$ & the projection matrices\\
$\kappa_i$ & eigenvalues of X-matrix \\
%\hline $$ & \\[1ex]
\hline
\end{tabular}
\end{table}

\section{3D nonlinear Stokes-Mueller equation}

The $3D$ nonlinear Stokes-Mueller polarimetry equation connects the Stokes vector of outgoing radiation with the nonlinear Stokes vector of incoming radiation and the nonlinear Mueller matrix of the material \cite{Shi}. The incoming fundamental  light has the frequencies of $\omega_1,...,\omega_n$ ($n=2$ for a three-wave mixing process such as SFG, and $n=3$ for a four-wave mixing (FWM) process) and the outgoing light generated by a particular nonlinear process possesses the frequency $\omega_\sigma$:
\begin{equation}\label{1}
  s'(\omega_\sigma)={\cal M}S(\omega_1,...,\omega_n).
\end{equation}
The $s'(\omega_\sigma)$ is a $9\times 1$ Stokes vector describing $3D$ polarization of outgoing radiation. The ${\cal M}$ is a $9\times 81$ Mueller matrix for the SFG process and the $S(\omega_1,\omega_2)$ is an $81\times 1$ nonlinear Stokes vector of incoming radiation.
For the FWM process, the ${\cal M}$ is a $9\times 9^3$ Mueller matrix and the $S(\omega_1,\omega_2,\omega_3)$ is a $9^3\times 1$ nonlinear Stokes vector of incoming radiation.

\subsection{3D Stokes Vector for outgoing radiation}

We generalize the Stokes vector of outgoing radiation for the case of $3D$ space.
Let us introduce the three component linear vector (analog of Jones's vector) by the relation
\begin{equation}\label{2}
 \Omega'=\left(
 \begin{array}{ccc}
 \tilde{E}'_x  \\
  \tilde{E}'_y  \\
  \tilde{E}'_z  \\
  \end{array}
 \right),
\end{equation}
where for nonlinear signal (for example SFG, or SHG for a pure state) $\tilde{E}'_i=E'_{0i}\exp[-i(\textbf{k}'\cdot \textbf{r}'-\omega_\sigma t-\varphi_i)]$ and $\textbf{r}'$ is the direction of the wave propagation of outgoing radiation. The prime denotes that the generalized Stokes vector is given for the nonlinearly generated light.
The coherency matrix for a pure state is given by the dyad product
\begin{equation}\label{3}
C=\Omega'\cdot \Omega'^\dag=\left(
\begin{array}{ccc}
     \tilde{E}'_x\tilde{E}'^{*}_x & \tilde{E}'_x\tilde{E}'^{*}_y  & \tilde{E}'_x\tilde{E}'^{*}_z  \\
\tilde{E}'_y\tilde{E}'^{*}_x  & \tilde{E}'_y\tilde{E}'^{*}_y  &\tilde{E}'_y\tilde{E}'^{*}_z  \\
   \tilde{E}'_z\tilde{E}'^{*}_x  & \tilde{E}'_z\tilde{E}'^{*}_y  & \tilde{E}'_z\tilde{E}'^{*}_z  \\
  \end{array}
  \right),
\end{equation}
where $\Omega'\cdot \Omega'^\dag$ is the matrix-diad and $\Omega'^\dag$ is a Hermitian conjugated vector. For a mixed state one has to use the time averaging matrix $\langle C\rangle$.

Now we introduce generalized Stokes vector in $3D$ space (see \cite{Brosseau})
\begin{equation}\label{4}
  s'_A=\mbox{Tr}(\langle C\rangle\lambda_A)=\langle\Omega'^\dag\lambda_A\Omega'\rangle,
\end{equation}
where $\lambda_A$ (index $A$ runs $0,1,...,8$) are Gell-Mann matrices \footnote{We changed the order of Gell-Mann matrices \cite{Kruglov}}
\[
\lambda_0=\sqrt{\frac{2}{3}}\left(
\begin{array}{ccc}
    1 & 0  & 0 \\
 0  & 1 & 0  \\
  0 & 0 & 1 \\
  \end{array}
  \right),~~~~ \lambda_1=\sqrt{\frac{1}{3}}\left(
\begin{array}{ccc}
    1 & 0  & 0 \\
 0  & 1 & 0  \\
  0 & 0 & -2 \\
  \end{array}
  \right),
 \]
 \[
  \lambda_2=\left(
\begin{array}{ccc}
    1 & 0  & 0 \\
 0  & -1 & 0  \\
  0 & 0 & 0 \\
  \end{array}
  \right),~~~~
 \lambda_3=\left(
\begin{array}{ccc}
    0 & 1  & 0 \\
 1 & 0 & 0  \\
  0 & 0 & 0 \\
  \end{array}
  \right),
\]
\[
 \lambda_4=\left(
\begin{array}{ccc}
    0 & 0  & 0 \\
 0  & 0 & 1  \\
  0 & 1 & 0 \\
  \end{array}
  \right),~~~ \lambda_5=\left(
\begin{array}{ccc}
    0 & 0  & 1 \\
 0  & 0 & 0  \\
  1 & 0 & 0 \\
  \end{array}
  \right),~~~
\]
\[
 \lambda_6=\left(
\begin{array}{ccc}
    0 & -i & 0 \\
 i & 0 & 0  \\
  0 & 0 & 0 \\
  \end{array}
  \right),~~~ \lambda_7=\left(
\begin{array}{ccc}
    0 & 0 & 0 \\
 0 & 0 & -i  \\
  0 & i & 0 \\
  \end{array}
  \right),
  \]
\begin{equation}\label{5}
  \lambda_8=\left(
\begin{array}{ccc}
    0 & 0 & -i \\
 0 & 0 & 0  \\
  i & 0 & 0 \\
  \end{array}
  \right),
\end{equation}
obeying the relations Tr$(\lambda_i\lambda_j)=2\delta_{ij}$, Tr$(\lambda_i)=0$ (if $i\neq 0)$.
Making use of Eqs. (3) and (4) we obtain the generalized Stokes vector
\begin{equation}\label{6}
  s'=\left(
 \begin{array}{ccc}
 s'_0  \\ s'_1 \\ s'_2  \\ s'_3  \\ s'_4 \\ s'_5  \\s'_6  \\ s'_7 \\ s'_8  \\
  \end{array}
 \right)=\langle\left(
 \begin{array}{ccc}
\sqrt{\frac{2}{3}}(\tilde{E}'_x\tilde{E}'^{*}_x +\tilde{E}'_y\tilde{E}'^{*}_y +\tilde{E}'_z\tilde{E}'^{*}_z ) \\
\sqrt{\frac{1}{3}}(\tilde{E}'_x\tilde{E}'^{*}_x +\tilde{E}'_y\tilde{E}'^{*}_y -2\tilde{E}'_z\tilde{E}'^{*}_z ) \\
\tilde{E}'_x\tilde{E}'^{*}_x -\tilde{E}'_y\tilde{E}'^{*}_y \\
\tilde{E}'_x\tilde{E}'^{*}_y +\tilde{E}'_y\tilde{E}'^{*}_x\\
\tilde{E}'_y\tilde{E}'^{*}_z +\tilde{E}'_z\tilde{E}'^{*}_y\\
\tilde{E}'_x\tilde{E}'^{*}_z +\tilde{E}'_z\tilde{E}'^{*}_x \\
i(\tilde{E}'_x\tilde{E}'^{*}_y -\tilde{E}'_y\tilde{E}'^{*}_x) \\
 i(\tilde{E}'_y\tilde{E}'^{*}_z -\tilde{E}'_z\tilde{E}'^{*}_y)  \\
 i( \tilde{E}'_x\tilde{E}'^{*}_z -\tilde{E}'_z\tilde{E}'^{*}_x)  \\
  \end{array}
 \right)\rangle.
\end{equation}
It can be noted that the first 3 elements of the $3D$ Stokes vector contain only quadratic terms of the electric fields. The $s'_3$, $s'_4$ and $s'_5$ components contain cross correlation terms between different electric field components, while $s'_6$, $s'_7$ and $s'_8$ components are related to the imaginary parts of the electric field components. In addition, $s'_3$ component is related to $s'_6$ component, while $s'_4$ is related to $s'_7$, and correspondingly $s'_5$ is related to $s'_8$ component. One can verify that for pure states, when we omit brackets $\langle...\rangle$ in Eq. (6), the relation $2s_0^{'2}=\sum_{N=1}^8s^{'2}_N$ holds. For impure states, $2s_0^{'2}>\sum_{N=1}^8s^{'2}_N$, so that
we can use the degree of polarization \cite{Kruglov} ${\cal P}=\sqrt{\sum_{N=1}^8s^{'2}_N/2s_0^{'2}}$ which ranges from $0$ to $1$ for unpolarized to fully polarized radiation, respectively. The rotational transformation of outgoing Stokes vector is given in Appendix C.

The measurement of $3D$ Stokes vector can be performed by determining the intensities of polarized light at $0^\circ$, $90^\circ$, $45^\circ$, $-45^\circ$, RCP, and LPC separately for the signal propagating in $x$, $y$ and $z$ directions of Cartesian laboratory reference frame
\begin{equation}\label{7}
 s'=
\left(
 \begin{array}{ccc}
\sqrt{\frac{1}{6}}\biggl(I_{yz}(0^\circ)+I_{yz}(90^\circ)+I_{xz}(0^\circ)+I_{xz}(90^\circ)\\
+I_{xy}(0^\circ)+I_{xy}(90^\circ) \biggr) \\
\sqrt{\frac{1}{3}}\left(I_{yz}(0^\circ)-I_{yz}(90^\circ)+I_{xz}(0^\circ)-I_{xz}(90^\circ) \right) \\
I_{xy}(0^\circ)-I_{xy}(90^\circ)  \\
I_{xy}(45^\circ)-I_{xy}(-45^\circ) \\
I_{yz}(45^\circ)-I_{yz}(-45^\circ) \\
I_{xz}(45^\circ)-I_{xz}(-45^\circ)  \\
I_{xy}(RCP)-I_{xy}(LCP) \\
I_{yz}(RCP)-I_{yz}(LCP)   \\
I_{xz}(RCP)-I_{xz}(LCP)  \\
  \end{array}
 \right).
\end{equation}
 The subscripts in Eq. (7) indicate the indexes of the normal planes with respect to the beam propagation directions.

\subsection{3D double Stokes vector of SFG}

The polarization vector ${\cal P}_i$ ($i=x, y, z$) contains the interacting incoming electric fields $\tilde{E}_i=\tilde{E}_{0i}\exp[-i(\textbf{k}\cdot \textbf{r}-\omega_i t-\phi_i)]$ ($i=1,2$) for a pure state, where $\textbf{r}$ is the direction of the wave propagation of incoming radiation. The polarization vector is given by
\begin{equation}\label{8}
 {\cal P}_i=\chi_{ijk}\Omega_j(\omega_1)\Omega_k(\omega_2)=\chi_{iA}\Psi_A(\omega_1,\omega_2),
\end{equation}
where $A=1,2,...,9$ and we use the contracted notation.
Here the state vector is defined as
\[
 \Psi(\omega_1,\omega_2)=\Omega(\omega_1)\otimes\Omega(\omega_2)
 \]
\[
 = \biggl(
 \tilde{E}_x(\omega_1)\tilde{E}_x(\omega_2), \tilde{E}_x(\omega_1)\tilde{E}_y(\omega_2), \tilde{E}_x(\omega_1)\tilde{E}_z(\omega_2),
 \]
 \[
 \tilde{E}_y(\omega_1)\tilde{E}_x(\omega_2), \tilde{E}_y(\omega_1)\tilde{E}_y(\omega_2), \tilde{E}_y(\omega_1)\tilde{E}_z(\omega_2),
 \]
\begin{equation}\label{9}
\tilde{E}_z(\omega_1)\tilde{E}_x(\omega_2), \tilde{E}_z(\omega_1)\tilde{E}_y(\omega_2), \tilde{E}_z(\omega_1)\tilde{E}_z(\omega_2)\biggr)^T,
\end{equation}
where $\otimes$ denotes the Kronecker product, and the superscript T denotes conjugate transposed.
The coherency matrix is given by $ \Psi(\omega_1,\omega_2) \cdot \Psi^\dagger(\omega_1,\omega_2)$ and is $9\times 9$-matrix. To introduce the double Stokes vector we need $81$ base matrices $\eta_N$ ($N=0,1,...,80$) with the properties Tr$(\eta_M\eta_N)=2\delta_{MN}$.
Let us introduce elements of entire matrix algebra $\varepsilon^{A,B}$ obeying the relations
for the product of matrices and matrix elements \cite{Kruglov4}
\begin{equation}\label{10}
 \varepsilon^{A,B}\varepsilon^{C,D}=\delta_{BC}\varepsilon^{A,D},~~~~(\varepsilon^{A,B})_{CD}=\delta_{AC}
 \delta_{BD},
\end{equation}
where indices $A$, $B$, $C$, and $D$ run $1,2,...,9$.
Thus, matrix $\varepsilon^{A,B}$ possesses only one nonzero element where row $A$ and coulomb $B$ cross. As the state vector $\Psi$ is $9$-component vector we need $81$ matrices $\eta_N$ with the dimension $9\times 9$ to express the double Stokes vector $S_N$ \cite{Kruglov}. Then matrices $\eta_N$ are given by (see also \cite{Kruglov})
\[
\eta_{0}=\frac{1}{\sqrt{4.5}}I,
\]
\[
\eta_{1}=\frac{1}{\sqrt{36}}\biggl(\varepsilon^{1.1}+
\varepsilon^{2,2}+\varepsilon^{3,3}+
\varepsilon^{4,4}+\varepsilon^{5,5}
\]
\[
+\varepsilon^{6,6}+\varepsilon^{7,7}+\varepsilon^{8,8}-8\varepsilon^{9,9}\biggr),
\]
\[
\eta_{2}=\frac{1}{\sqrt{28}}\biggl(\varepsilon^{1.1}
+\varepsilon^{2,2}+\varepsilon^{3,3}
\]
\[
+\varepsilon^{4,4}+\varepsilon^{5,5}+\varepsilon^{6,6}+\varepsilon^{7,7}-7\varepsilon^{8,8}\biggr),
\]
\[
\eta_{3}=\frac{1}{\sqrt{21}}\left(\varepsilon^{1.1}+
\varepsilon^{2,2}+\varepsilon^{3,3}+
\varepsilon^{4,4}+\varepsilon^{5,5}++\varepsilon^{6,6}-6\varepsilon^{7,7}\right),
\]
\[
\eta_{4}=\frac{1}{\sqrt{15}}\left(\varepsilon^{1.1}+
\varepsilon^{2,2}+\varepsilon^{3,3}+
\varepsilon^{4,4}+\varepsilon^{5,5}-5\varepsilon^{6,6}\right),
\]
\[
\eta_{5}=\frac{1}{\sqrt{10}}\left(\varepsilon^{1.1}+\varepsilon^{2,2}+\varepsilon^{3,3}+
\varepsilon^{4,4}-4\varepsilon^{5,5}\right),
\]
\[
\eta_{6}=\frac{1}{\sqrt{6}}\left(\varepsilon^{1.1}+\varepsilon^{2,2}+\varepsilon^{3,3}-
3\varepsilon^{4,4}\right),
\]
\[
 \eta_{7}=\frac{1}{\sqrt{3}}\left(\varepsilon^{1.1}+\varepsilon^{2,2}-2\varepsilon^{3,3}\right),~
\eta_{8}=\varepsilon^{1.1}-\varepsilon^{2,2},
\]
\[
\eta_{E}=\varepsilon^{A,B}+\varepsilon^{B,A}~(A,B=1,...,9,~~E=9,...,44)~(A<B),
\]
\begin{equation}\label{11}
\eta_{F}=i(\varepsilon^{C,D}-\varepsilon^{D,C})~(C,D=1,...,9,~~F=45,...,80)~(C<D),
\end{equation}
where $I$ is the unit (identity) matrix, $\eta_{0},...\eta_{8}$ are diagonal matrices, and we imply the summation on repeated indices. $\eta_9,...\eta_{44}$ are symmetric matrices, $\eta_{45},...\eta_{80}$ are antisymmetric matrices, and Tr($\eta_N)=0$ (if $N\neq 0$). These matrices $\eta_N$ can be considered as generators of $SU(9)$ group. The state vector $ \Psi(\omega_1,\omega_2)$ under rotation transforms to $\Psi_R=(R\otimes R)\Psi$ where $R$ is $3\times 3$ rotation matrix (see Appendix C).

The intensity of SFG is
\begin{equation}\label{12}
  I\propto |{\cal P}_i|^2=\chi_{iA}\chi_{iB}^*\Psi_A\Psi^*_B.
\end{equation}
The coherency matrix of incoming light for pure state is a dyad and reads
\begin{equation}\label{13}
 \rho=\Psi \cdot \Psi^\dag.
\end{equation}
Thus, we imply that incoming light is polarized, i.e. the state is pure.
The double Stokes vector can be represented in the form
\begin{equation}\label{14}
  S_N=\mbox{Tr}(\rho \eta_N)=\Psi^\dag\eta_N\Psi,
\end{equation}
where matrixes $\eta_N$ ($N=0,1,2,...80$) are given by Eq. (11). The number of components can be contracted to 36 for a degenerate double Stokes vector of SHG presented in Appendix B. However, for a rotational covariant coherency matrix $\langle C\rangle$ a full bases with 81 components has to be kept.

\subsection{$3D$ nonlinear Mueller matrix}

By virtue of Eqs. (2) and (14), Eq. (1) can be written as
\begin{equation}\label{15}
  \Omega'^\dag\lambda_\alpha \Omega'={\cal M}_{\alpha N}\Psi^\dag\eta_N\Psi.
\end{equation}
The electric field of outgoing radiation is proportional to the polarization vector, $\tilde{E}'_i=A{\cal P}_i$. Then Eq. (2) becomes
\begin{equation}\label{16}
   \Omega'=A\left(
 \begin{array}{ccc}
 \chi_{xB}\Psi_B  \\
  \chi_{yB}\Psi_B \\
  \chi_{zB}\Psi_B  \\
  \end{array}
 \right).
\end{equation}
Placing Eq. (16) into (15) we obtain
\begin{equation}\label{17}
  A^2\chi^*_{iB}\Psi^*_B(\lambda_\alpha)_{ij}\chi_{jA}\Psi_A= {\cal M}_{\alpha N}\Psi^*_B(\eta_N)_{BA}\Psi_A.
\end{equation}
Because Eq. (17) holds for any functions $\Psi^*_B$, $\Psi_A$, we can omit them and write
\begin{equation}\label{18}
  A^2\chi^*_{iB}(\lambda_\alpha)_{ij}\chi_{jA}= {\cal M}_{\alpha N}(\eta_N)_{BA}.
\end{equation}
By multiplying Eq. (18) by $(\eta_M)_{AB}$ and using the relation Tr$(\eta_N\eta_M)=2\delta_{NM}$ one finds
\begin{equation}\label{19}
{\cal M}_{\alpha N}=\frac{1}{2} A^2\chi^*_{iB}(\lambda_\alpha)_{ij}\chi_{jA}(\eta_N)_{AB}.
\end{equation}
Eq. (19) can be rewritten in the matrix form
\begin{equation}\label{20}
  {\cal M}_{\alpha N}=\frac{1}{2} A^2\mbox{Tr}(\lambda_\alpha\chi\eta_N\chi^\dag).
\end{equation}
For simplicity we omit below the proportionality factor $A^2$. The Mueller matrix characterizes the nonlinear properties of the material. Knowing the ${\cal M}_{\alpha N}$ from experiments one can extract the products $\chi_{aA}\chi_{bB}^*$ from Eq. (20).
Thus, the ultrastructure of the material can be studied.

The Mueller matrix components can be obtained by measuring the intensities of outgoing radiation. One can prepare the pure polarization states of incoming radiation using polarization state generators (PSG) of laser beams incoming from perpendicular directions. The outgoing radiation polarization can be measured by a polarization state analysers (PSA) positioned at perpendicular directions in the laboratory coordinate system. Double Mueller matrix ${\cal M}_{\alpha N}$ can be found by making use of $81$ fundamental radiation polarization states ($Q=0,1,2,...,80$) defined by the double Stokes vector elements $S_{N,Q}$. For measuring the outgoing radiation Stokes vector components $s'_{\alpha,Q}$ we may write the equation as follows:
\begin{equation}\label{21}
  s'_{\alpha,Q}={\cal M}_{\alpha N}S_{N,Q},
\end{equation}
where SFG process requires to measure $s'_{\alpha,Q}$ matrix containing $9\times 81$ elements and prepare 81 polarization states resulting in $S_{N,Q}$ matrix of $81\times 81$ elements. We require that incoming
polarization states to be chosen in such a way that matrix $S_{N,Q}$ is invertible, i.e. the matrix $S^{-1}_{N,Q}$ exists. Then from Eq. (21) one can obtain the nonlinear Mueller matrix ${\cal M}_{\alpha N}=s'_{\alpha,Q}S^{-1}_{N,Q}$.
To prepare the double Stokes polarization state one can explore PSG that includes a quarter and a half wave plates for the beams pointing in orthogonal directions. One can find double Stokes polarization states expressed via incoming linear $3D$ Stokes parameters $s_A$ ($A=0,1,...,8$).
This allows us to obtain $S_{N,Q}$ and from the measurements to find $s'_{\alpha,Q}$. Then from Eq. (21) one may calculate the Mueller matrix ${\cal M}_{\alpha,N}$. For the degenerate process of SHG the double Stokes vector assumes contracted form presented in Appendix B. However, a full double Stokes vector is required, and can be constructed from contracted SHG polarimetry measurements, for obtaining rotationally covariant $X$-matrix presented in the next section.

\section{The susceptibility correlation matrix $X$}

The $X$-matrix can be constructed from the Mueller matrix ${\cal M}$ by the relation \cite{Kruglov1}
\begin{equation}\label{22}
X=\frac{1}{2} {\cal M}_{tN}F_{tN},
\end{equation}
where $F_{tN}=\lambda_t\otimes \eta_N^T$. Equation (22) allows to transform Mueller matrix ${\cal M}$ into square matrix $X$ for the SFG process. It is convenient to analyze the coherency-like Hermitian matrix $X$ because it possesses the positive eigenvalues $\kappa_1$,...,$\kappa_n$
 ($27$ eigenvalues for SFG). In this case we have $81$ matrices $\eta_N$ ($N=0,1,...,80$), and the $9\times 81$ Mueller matrix can be transformed into square $27\times 27$ $X$-matrix. For this case ($n=27$), the rotational invariance of eigenvalues of the $X$-matrix is guaranteed (see Appendix C).
$X$-matrix can be represented as follows \cite{Kruglov1}:
\begin{equation}\label{23}
	X=\sum_i^{n}\kappa_iP_i,
\end{equation}
where $\kappa_i$ are eigenvalues of $X$-matrix, and $P_i$ are projection matrices with properties
$P_i^2=P_i$, $P_iP_j=0$ ($i\neq j$). Then the trace of the coherency matrix $X$ can be expressed as:
\begin{equation}\label{24}
\mbox{Tr}(X)=\kappa_1+...+\kappa_n.
\end{equation}
The eigenvalues can be used to estimate the purity of the $X$-matrix.

\subsection{Index of depolarization of $X$-matrix}

The heterogeneous media will provide an averaged response containing multiple pure states. The heterogeneity of $X$-matrix can be characterized by the index of depolarization, which is derived as follows. For any eigenvalues of the $X$-matrix the following inequalities can be defined:
\begin{equation}\label{25}
  n(\kappa_1^2+\kappa_2^2+...+\kappa_n^2)\geq (\kappa_1+\kappa_2+...+\kappa_n)^2\geq \kappa_1^2+\kappa_2^2+...+\kappa_n^2.
\end{equation}
Let us consider the Frobenius norm that is defined by
\begin{equation}\label{26}
  \|X\|_F=\left[\mbox{Tr}\left(X^\dag X\right)\right]^{1/2}.
\end{equation}
The squared Frobenius norm can be calculated as follows:
\begin{equation}\label{27}
\|X\|_F^2=\kappa_1^2+\kappa_2^2+...+\kappa_n^2.
\end{equation}
Taking into account that Tr$(X)=\kappa_1+\kappa_2+...+\kappa_n$ we can rewrite Eq. (25) as follows:
\begin{equation}\label{28}
  n(\mbox{Tr}(X))^2\geq n\|X\|_F^2\geq (\mbox{Tr}(X))^2,
\end{equation}
or in the equivalent form
\begin{equation}\label{29}
 1\geq\frac{n\|X\|_F^2-(\mbox{Tr}(X))^2}{(n-1)(\mbox{Tr}(X))^2} \geq 0.
\end{equation}
Then we can introduce the index of depolarization (see also \cite{Gil2})
\begin{equation}\label{30}
P_D=\sqrt{\frac{1}{n-1}\left(n\left[\frac{\|X\|_F}{\mbox{Tr}(X)}\right]^2-1\right)},
\end{equation}
so that $1\geq P_D\geq 0$.
For the linear process when the Mueller matrix is $4\times4$ matrix ($n=4$) we come to the definition introduced in \cite{Gil1} (see also \cite{Gil2}, \cite{Roy}).
We can verify that the index of depolarization $P_D=0$ for the totally depolarizing matrix $X$, and $P_D=1$ for the pure polarized state. For the totally depolarizing matrix
\begin{equation}\label{31}
  X=aI,~~~ \|X\|_F=a\sqrt{n},~~~\mbox{Tr}(X)=an,
\end{equation}
where $a$ is arbitrary parameter and $I$ is the identity matrix. One can check that for this case $P_D=0$.
For the pure polarized state
\begin{equation}\label{32}
  X=\kappa P,~~~ \|X\|_F=\kappa,~~~\mbox{Tr}(X)=\kappa,
\end{equation}
where $P$ is the projection matrix, $P^2=P$ and, as a result, $P_D=1$. Therefore, $P_D$ parameter characterizes disorder in the material, which is invariant under rotation (see Appendix C for consideration of rotational invariance).

\subsection{Decomposition of $X$-matrix into two pure polarization states}

The $X$-matrix reflects properties of the media, therefore various decompositions can be designed to model the underlying structural organization of the material. If we consider two pure polarization states present in the material, the $X$-matrix can be expressed as follows:
\begin{equation}\label{33}
X=\kappa_1P_1+\kappa_2P_2.
\end{equation}
 where $P_1$, $P_2$ are the projection matrices, $P_1^2=P_1$, $P_2^2=P_2$, $P_1P_2=0$.
In this situation we have
\begin{equation}\label{34}
\|X\|_F=\sqrt{\kappa_1^2+\kappa_2^2},~~~\mbox{Tr}(X)=\kappa_1+\kappa_2,
\end{equation}
and
\begin{equation}\label{35}
P_D=\frac{\sqrt{2(\kappa_1^2+\kappa_2^2)-(\kappa_1+\kappa_2)^2}}{(\kappa_1+\kappa_2)}=
\frac{\kappa_1-\kappa_2}{\kappa_1+\kappa_2}.
\end{equation}
We imply here that $\kappa_1\geq \kappa_2$.
Depending on the relative contributions of the two pure polarization states, the index of depolarization will vary from almost purely polarized state to fully unpolarized state. The graph of index of polarization dependence on the ratio of $\kappa_1/\kappa_2$ is presented in Fig. 1.
\begin{figure}[h]
\includegraphics[height=3.0in,width=3.0in]{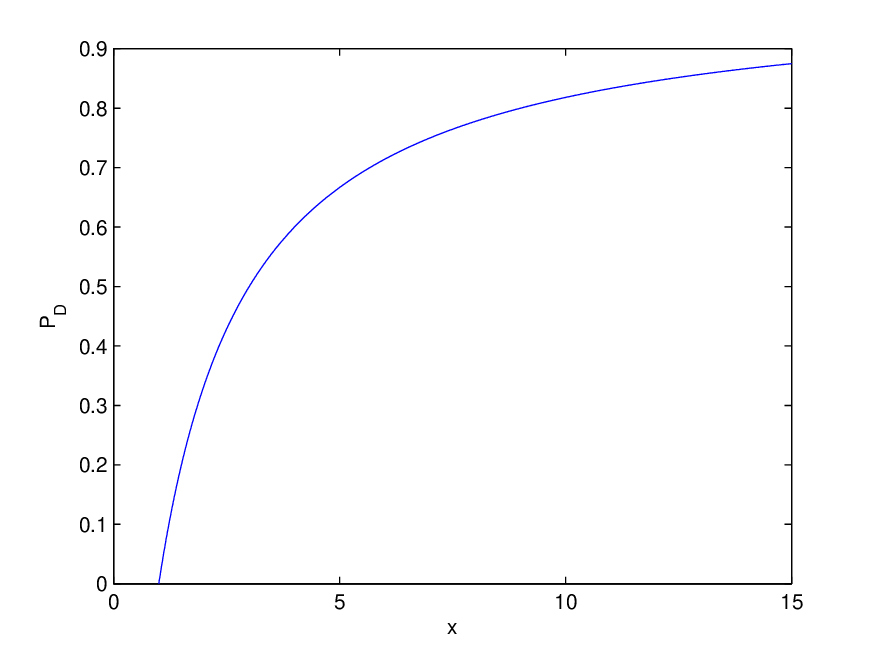}
\caption{\label{fig.1}The plot of the function $P_D$ vs. $x=\kappa_1/\kappa_2$.}
\end{figure}

\subsection{$X$-matrix as a sum of depolarizing matrix and pure polarized state}

The polarization measurements often have a noise contribution, therefore the $X$-matrix can be represented as a superposition of totally depolarizing matrix and a matrix for a pure polarization state:
\begin{equation}\label{36}
X=\kappa_bI+\kappa P,
\end{equation}
where $I$ is identity matrix, $P^2=P$.
Then:
\[
\|X\|_F=\sqrt{\kappa(\kappa+2\kappa_b)+\kappa_b^2n},
\]
\begin{equation}\label{37}
  \mbox{Tr}(X)=\kappa+\kappa_bn,~P_D=\frac{\kappa}{\kappa+\kappa_bn}.
\end{equation}
From Eqs. (24) and (37) we obtain the equation
\begin{equation}\label{38}
\sum_i^{n}\kappa_iP_i=\kappa_bI+\kappa P.
\end{equation}
Squaring Eq. (38) one finds
\begin{equation}\label{39}
\sum_i^{n}\kappa_i^2P_i=\kappa_b^2I+\kappa_b\kappa P+\kappa^2 P.
\end{equation}
Making trace of left and right sides of Eqs. (38) and (39) we obtain the system of two equations
\begin{equation}\label{40}
\sum_i^{n}\kappa_i=n\kappa_b+\kappa,~~~~ \sum_i^{n}\kappa_i^2=n\kappa_b^2+2\kappa_b\kappa+\kappa^2.
\end{equation}
Solving the system of two equations (40) one finds
\begin{equation}\label{41}
\kappa_b=\frac{\sqrt{n-1}\sum_i^{n}\kappa_i-\sqrt{n\sum_i^{n}\kappa_i^2-
		(\sum_i^{n}\kappa_i)^2}}{n\sqrt{n-1}}.
\end{equation}
In Eq. (41) we wrote down only one root to have positive eigenvalues $\kappa_b$ and $\kappa$.
Making use of Eqs. (37) and (41) we find
\begin{equation}\label{42}
\kappa_b=\frac{(1-P_D)\sum_i^{n}\kappa_i}{n}.
\end{equation}
From Eq. (40) one obtains
\begin{equation}\label{43}
\kappa=P_D\sum_i^{n}\kappa_i.
\end{equation}
It can be seen that $\kappa_b$ and $\kappa$ can be expressed via the index of depolarization and the sum of the eigenvalues of the $X$-matrix. If we have only two eigenvalues $\kappa_1$ and $\kappa_2$ of the $X$-matrix, by plugging index of depolarization expression Eq. (35) into Eqs. (42) and (43) one recovers that $\kappa_b=\kappa_2$ and $\kappa=\kappa_1-\kappa_2$.
Thus, the decomposition of $X$-matrix is not unique. Different sources of SFG can reproduce the same coherency-like matrix $X$. The same situation takes place in the linear $2D$ case when we have $4\times 4$ coherency matrix.

\section{Conclusion}

In this paper we have introduced $3D$ linear and nonlinear Stokes vectors and considered a special case of SFG polarimetry. The $3D$ Stokes vector of SFG radiation was derived and a parametrization of the coherency matrix possessing $9$ elements was briefly considered. We have derived the expression for the double Stokes vector $S$ for the pure state via generalized $3D$ Stokes vector $s$ of the broadband laser radiation. The elements of $9\times 9$ nonlinear coherency matrix of interacting fundamental radiation electric fields were expressed respectively for SFG and SHG. For SHG, considered in Appendix B, we have used a contracted 6-component state vector $\psi$. In this case eigenvalues of the $X$-matrix are not rotational invariant. To have the rotational covariant $X$-matrix one should use 9-component state vector as for SFG.
% In this case, to transfer from SFG to SHG, we need to put frequencies $\omega_1=\omega_2$ for the coherence matrix and other expressions of %SFG.
We have derived double Mueller matrix, which is expressed via susceptibility tensor elements. It is described how to calculate the Mueller matrix elements from the polarimetric measurements. This allows to study the molecular organization of different materials. The coherency-like matrix $X$ for SFG is introduced. Thus, $9\times 81$ Mueller matrix is transformed into $27\times 27$
square matrix $X$. For the degenerate case of SHG the Mueller matrix is transformed into $18\times 18$ square matrix $X$, while FWM $9\times 9^3$ Mueller matrix is transformed to $81\times 81$ square matrix $X$. It should be noted that for the degenerate cases $X$-matrix is not rotationally covariant.
%However, for the degenerate cases the eigenvalues of the $X$-matrices are not rotational invariant values (see Appendix C).

The $X$-matrix is Hermitian semi-definite positive matrix and, therefore, its eigenvalues are non-negative. If some eigenvalues obtained from measurements are negative they are nonphysical and should be set to zero. The same procedure of filtering (''quick and dirty") is used in quantum tomography \cite{James}, \cite{James1}.
Another method of most likely estimation (MLE) allows us to obtain the most probable $X$-matrix
from the measured Stokes matrix without calculating the Mueller matrix. MLE, which is less susceptible to experimental noise, can be performed by Cholesky decomposition \cite{Cholesky}.
For the case of one nonzero eigenvalue of $X$-matrix, it represents the pure state and the susceptibility can be evaluated. If there are several positive eigenvalues of matrix $X$, they correspond to different sources with their own susceptibilities. An alternative interpretation is also given for the $X$-matrix as a composition of the isotropic background and pure state with one eigenvalue. Different decompositions of $X$-matrix can be applied to investigate the underlying organization of the material. The decompositions are not unique and therefore assumptions are required to choose the most probable decomposition model. The index of depolarization for $X$-matrix is introduced ranging from $P_D=0$ for the totally depolarizing matrix ${\cal M}$ and $P_D=1$ for the pure polarized state. The theory of $3D$ nonlinear polarimetry provides a basis for the development of novel polarimetric measurement methods and innovative experimental set ups. The $3D$ nonlinear Stokes-Mueller  polarimetry is developed with possible applications in mind for the ultrastructural investigations of biological specimens with nonlinear polarimetric microscopy.

\section{Appendix A: The coherency matrix}

Making use of Eq. (6) for outgoing radiation, we obtain the relations
\[
\langle\tilde{E'}_x\tilde{E}'^{*}_x\rangle=\frac{1}{2\sqrt{3}}s'_1+\frac{1}{2}s'_2+\frac{1}{\sqrt{6}}s'_0,
\]
\[
\langle\tilde{E'}_y\tilde{E}'^{*}_y\rangle=\frac{1}{2\sqrt{3}}s'_1-\frac{1}{2}s'_2+\frac{1}{\sqrt{6}}s'_0,
\]
\[
\langle\tilde{E'}_z\tilde{E}'^{*}_z\rangle=\frac{1}{\sqrt{6}}s'_0-\frac{1}{\sqrt{3}}s'_1,
\]
\[
\langle\tilde{E'}_x\tilde{E}'^{*}_y\rangle=\frac{1}{2}(s'_3-is'_6),~~~
\langle\tilde{E'}_y\tilde{E}'^{*}_x\rangle=\frac{1}{2}(s'_3+is'_6),
\]
\[
\langle\tilde{E'}_y\tilde{E}'^{*}_z\rangle=\frac{1}{2}(s'_4-is'_7),~~~
\langle \tilde{E'}_z\tilde{E}'^{*}_y\rangle=\frac{1}{2}(s'_4+is'_7),
\]
\begin{equation}\label{44}
\langle\tilde{E'}_x\tilde{E}'^{*}_z\rangle=\frac{1}{2}(s'_5-is'_8),~~~
\langle\tilde{E'}_z\tilde{E}'^{*}_x\rangle=\frac{1}{2}(s'_5+is'_8).
\end{equation}
%The parametrization of the coherency matrix $\langle C\rangle$ for the fundamental radiation is given in Appendix A.
Similar relations hold for incoming radiation (for letters without primes). In that case, electric fields with different frequencies have to be considered for SFG.
%Then from Eqs. (9), (11) and (14) one can find the $3D$ double Stokes vector $S$ expressed through the generalized fundamental radiation Stokes %vectors.
%The nonlinear Stoke vectors for non-degenerate higher order processes can be constructed by choosing appropriate orthogonal base matrices.

From Eqs. (3) and (44) we obtain
%\begin{widetext}
\begin{equation}\label{45}
\langle C\rangle=\left(
\begin{array}{ccc}
\frac{1}{2}\left(\frac{1}{\sqrt{3}}s'_1+s'_2+\sqrt{\frac{2}{3}}s'_0 \right)&\frac{1}{2}(s'_3-is'_6)&\frac{1}{2}(s'_5-is'_8)\\
\frac{1}{2}(s'_3+is'_6)&\frac{1}{2}\left(\frac{1}{\sqrt{3}}s'_1-s'_2+\sqrt{\frac{2}{3}}s'_0 \right)&\frac{1}{2}(s'_4-is'_7)\\
\frac{1}{2}(s'_5+is'_8)&\frac{1}{2}(s'_4+is'_7)&\frac{1}{\sqrt{3}}\left(\frac{1}{\sqrt{2}}s'_0-s'_1\right)\\
  \end{array}
  \right)
\end{equation}
%\end{widetext}
For pure polarized light the nine Stokes parameters $s'_i$ are not independent and there are relationships between them. So, there are only five independent parameters \cite{Sheppard}.
One can parameterize matrix $\langle C\rangle$, possessing nine parameters, by \cite{Dennis}, \cite{Gil}
\[
\langle C\rangle=O\langle J\rangle O^T,~~~~\langle J\rangle=A+iN,
\]
\begin{equation}\label{46}
A=
  \left(
\begin{array}{ccc}
a_1 & 0 & 0 \\
0 & a_2 & 0 \\
0 & 0 & a_3 \\
  \end{array}
  \right),~~~~
  N=\left(
\begin{array}{ccc}
0 & -n_3 & n_2 \\
n_3 & 0 & -n_1 \\
-n_2 & n_1 & 0 \\
  \end{array}
  \right),
\end{equation}
where the rotation matrix $O$ has three parameters (Euler's angles or Gibbs's parameters \cite{Kruglov5}), $a_1$, $a_2$, $a_3$ correspond to the coherency matrix which is the superposition of three linearly polarized pure states \cite{Gil}, and $n_1$, $n_2$, $n_3$ are the components of
the angular momentum of the wave in new reference frame. Making use of Eq. (6), one can verify that the axial vector $\textbf{n}=(n_1,n_2,n_3)$, where $n_1=s'_7$, $n_2=-s'_8$,
$n_3=s'_6$, satisfies the equation $\tilde{\textbf{E}}'\cdot \textbf{n}=0$, and $\textbf{n}$ describes circular polarization \cite{Carozzi}. The electric field $\tilde{\textbf{E}}'$ belongs to the plane $\textbf{n}\cdot\textbf{r}=0$ ($\textbf{r}=(x,y,z)$) while the direction of the axial vector $\textbf{n}$ is defined by the equation $x/n_1=y/n_2=z/n_3$ \cite{Sheppard}. The matrix $N$ can be represented as $N=(N_{ij})$ with matrix elements $N_{ij}=-\epsilon _{ijk}n_k$ (we imply a summation over repeated indices), where $\epsilon _{ijk}$ is antisymmetric Levy-Civita symbol.
The characteristic equation det($N-\kappa I)=0$ ($I$ is an identity matrix) reads $\kappa^3+\kappa(n_1^2+n_2^2+n_3^2)=0$ so that eigenvalues of the matrix $N$ are, $\kappa_1=0$, $\kappa_2=\sqrt{\textbf{n}^2}$, $\kappa_3=-\sqrt{\textbf{n}^2}$.
This parametrization can be used in experiments with non-collinear beam geometries, highly converging/diverging beams,
or where there is not well-defined propagation direction of light.

\section{Appendix B: SHG Double Stokes vector}

In the degenerate case of SFG, i.e. SHG, the number of double Stokes vector components can be contracted. It should be noted, however, that by contracting the dimension of the state vector we lose the rotational covariance of the coherence matrix (see Appendix C). However, the full double Stokes vector can be restored by complementing the degenerate Stokes components values obtained in SHG.
For the SHG generation ($\omega_1=\omega_2$) $\chi^{(2)}_{ijk}\equiv \chi^{(2)}_{iA}$ ($A=xx, yy, zz, xy, xz, yz$) and we use the contracted notation \cite{Boyd}. Then the state vector is defined as follows:
\begin{equation}\label{47}
  \psi=\left (
 \begin{array}{cccccc}
 \tilde{E}^2_x  \\
  \tilde{E}^2_y  \\
  \tilde{E}^2_z  \\
  2\tilde{E}_x \tilde{E}_y \\
  2\tilde{E}_x\tilde{E}_z  \\
  2\tilde{E}_y\tilde{E}_z  \\
  \end{array}
  \right).
\end{equation}
Thus, for simplicity, we use 6-component state vector. The coherency matrix of incoming light for pure state is a dyad and reads
%\begin{widetext}
\[
\rho=\psi \cdot \psi^\dag
\]
\begin{equation}\label{48}
=
  \left(
\begin{array}{cccccc}
   |\tilde{E}_x|^4  & \tilde{E}_x^2 \tilde{E}_y^{*2}  & \tilde{E}_x^2 \tilde{E}_z^{*2}&2\tilde{E}_x^2 \tilde{E}_x^{*}\tilde{E}_y^{*} & 2\tilde{E}_x^2 \tilde{E}_x^{*}\tilde{E}_z^{*}& 2\tilde{E}_x^2 \tilde{E}_y^{*}\tilde{E}_z^{*} \\
 \tilde{E}_y^2 \tilde{E}_x^{*2} & |\tilde{E}_y|^4  & \tilde{E}_y^2 \tilde{E}_z^{*2}&  2\tilde{E}_y^2 \tilde{E}_x^{*}\tilde{E}_y^{*}&  2\tilde{E}_y^2 \tilde{E}_x^{*}\tilde{E}_z^{*} & 2\tilde{E}_y^2 \tilde{E}_y^{*}\tilde{E}_z^{*}  \\
   \tilde{E}_z^2 \tilde{E}_x^{*2} & \tilde{E}_z^2 \tilde{E}_y^{*2}  & |\tilde{E}_z|^4&  2\tilde{E}_z^2 \tilde{E}_x^{*}\tilde{E}_y^{*}& 2\tilde{E}_z^2 \tilde{E}_x^{*}\tilde{E}_z^{*} & 2\tilde{E}_z^2 \tilde{E}_y^{*}\tilde{E}_z^{*} \\
   2\tilde{E}_x\tilde{E}_y \tilde{E}_x^{*2} & 2\tilde{E}_x\tilde{E}_y \tilde{E}_y^{*2}  & 2\tilde{E}_x\tilde{E}_y \tilde{E}_z^{*2}&4|\tilde{E}_x|^2|\tilde{E}_y|^2 &  4|\tilde{E}_x|^2\tilde{E}_y \tilde{E}_z^{*} & 4\tilde{E}_x|\tilde{E}_y|^2 \tilde{E}_z^{*} \\
   2\tilde{E}_x\tilde{E}_z \tilde{E}_x^{*2} & 2\tilde{E}_x\tilde{E}_z \tilde{E}_y^{*2}  & 2\tilde{E}_x\tilde{E}_z \tilde{E}_z^{*2}& 4|\tilde{E}_x|^2\tilde{E}_z \tilde{E}_y^{*} & 4|\tilde{E}_x|^2|\tilde{E}_z|^2  & 4\tilde{E}_x|\tilde{E}_z|^2\tilde{E}_y^{*} \\      2\tilde{E}_y\tilde{E}_z \tilde{E}_x^{*2} & 2\tilde{E}_y\tilde{E}_z \tilde{E}_y^{*2}  & 2\tilde{E}_y\tilde{E}_z \tilde{E}_z^{*2} &
   4|\tilde{E}_y|^2\tilde{E}_z \tilde{E}_x^{*} &   4\tilde{E}_y|\tilde{E}_z|^2 \tilde{E}_x^{*} & 4|\tilde{E}_y|^2|\tilde{E}_z|^2\\
    \end{array}
 \right).
\end{equation}
Thus, we imply that incoming light is polarized, i.e. the state is pure. The double Stokes vector can be represented in the form:
\begin{equation}\label{49}
  S_N=\mbox{Tr}(\rho \eta_N)=\psi^\dag\eta_N\psi,
\end{equation}
where $\eta_N$ ($N=0,1,...35$) are $6\times 6$ matrices with the relations Tr$(\eta_M\eta_N)=2\delta_{MN}$.
The matrices $\eta_N$ here are given by
\[
\eta_{0}=\frac{1}{\sqrt{3}}I,~~~\eta_{1}=\frac{1}{\sqrt{15}}\left(\varepsilon^{1.1}+
\varepsilon^{2,2}+\varepsilon^{3,3}+
\varepsilon^{4,4}+\varepsilon^{5,5}-5\varepsilon^{6,6}\right),
\]
\[
\eta_{2}=\frac{1}{\sqrt{10}}\left(\varepsilon^{1.1}+\varepsilon^{2,2}+\varepsilon^{3,3}+
\varepsilon^{4,4}-4\varepsilon^{5,5}\right),
\]
\[
\eta_{3}=\frac{1}{\sqrt{6}}\left(\varepsilon^{1.1}+\varepsilon^{2,2}+\varepsilon^{3,3}-
3\varepsilon^{4,4}\right),
\]
\[
 \eta_{4}=\frac{1}{\sqrt{3}}\left(\varepsilon^{1.1}+\varepsilon^{2,2}-2\varepsilon^{3,3}\right),~
\eta_{5}=\varepsilon^{1.1}-\varepsilon^{2,2},
\]
\[
\eta_{E}=\varepsilon^{A,B}+\varepsilon^{B,A}~~(A,B=1,...,6,~~E=6,...,20)~(A<B),
\]
\begin{equation}\label{50}
\eta_{F}=i(\varepsilon^{C,D}-\varepsilon^{D,C})~(C,D=1,...,6,~~F=21,...,35\emph{})~(C<D).
\end{equation}
Here $\eta_{0},...\eta_{5}$ are diagonal matrices, $\eta_6,...\eta_{20}$ are symmetric matrices, $\eta_{21},...\eta_{35}$ are antisymmetric matrices. The matrices $\eta_N$ are the generators of $SU(6)$ group.

The parametrization of the coherency matrix $\langle C\rangle$ is given in Appendix A. Similar relations hold for incoming radiation (letters without primes).

From Eqs. (47), (49) and (50) we obtain double Stokes vector
\[
 \left(
 \begin{array}{ccccccccc}
S_0  \\
S_1 \\
S_2 \\
S_3 \\
S_4\\
S_5\\
S_6\\
S_7\\
S_8\\
  \end{array}
 \right) =
\left(
  \begin{array}{cccccccccccc}
 \frac{1}{\sqrt{3}}(|\tilde{E}_x|^4+|\tilde{E}_y|^4+|\tilde{E}_z|^4+4|\tilde{E}_x|^2|\tilde{E}_y|^2\\
 +4|\tilde{E}_x|^2|\tilde{E}_z|^2+4|\tilde{E}_z|^2|\tilde{E}_y|^2)\\
\frac{1}{\sqrt{15}} (|\tilde{E}_x|^4+|\tilde{E}_y|^4+|\tilde{E}_z|^4+4|\tilde{E}_x|^2|\tilde{E}_y|^2\\
+4|\tilde{E}_x|^2|\tilde{E}_z|^2-20|\tilde{E}_z|^2|\tilde{E}_y|^2)\\
\frac{1}{\sqrt{10}}(|\tilde{E}_x|^4+|\tilde{E}_y|^4+|\tilde{E}_z|^4+4|\tilde{E}_x|^2|\tilde{E}_y|^2\\
-16|\tilde{E}_x|^2|\tilde{E}_z|^2)\\
\frac{1}{\sqrt{6}}(|\tilde{E}_x|^4+|\tilde{E}_y|^4+|\tilde{E}_z|^4-12|\tilde{E}_x|^2|\tilde{E}_y|^2)\\
\frac{1}{\sqrt{3}}(|\tilde{E}_x|^4+|\tilde{E}_y|^4-2|\tilde{E}_z|^4)\\
|\tilde{E}_x|^4-|\tilde{E}_y|^4\\
\tilde{E}_x^2\tilde{E}_y^{*2}+\tilde{E}_y^2\tilde{E}_x^{*2}\\
\tilde{E}_x^2\tilde{E}_z^{*2}+\tilde{E}_z^2\tilde{E}_x^{*2}\\
2(\tilde{E}_x^2\tilde{E}_y^{*}\tilde{E}_x^*+ \tilde{E}_x\tilde{E}_y\tilde{E}_x^{*2})\\
  \end{array}
 \right),
 \]
 \[
   \left(
 \begin{array}{ccccccccc}
S_{9}  \\
S_{10} \\
S_{11} \\
S_{12} \\
S_{13}\\
S_{14}\\
S_{15}\\
S_{16}\\
S_{17}\\
  \end{array}
 \right)= \left(
  \begin{array}{ccccccccc}
  2(\tilde{E}_x^2\tilde{E}_z^{*}\tilde{E}_x^*+ \tilde{E}_x\tilde{E}_z\tilde{E}_x^{*2}) \\
2(\tilde{E}_x^2\tilde{E}_z^{*}\tilde{E}_y^*+ \tilde{E}_y\tilde{E}_z\tilde{E}_x^{*2})\\
\tilde{E}_z^2\tilde{E}_y^{*2}+ \tilde{E}_y^2\tilde{E}_z^{*2}\\
2(\tilde{E}_y^2\tilde{E}_x^{*}\tilde{E}_y^*+ \tilde{E}_y\tilde{E}_x\tilde{E}_y^{*2})\\
2(\tilde{E}_y^2\tilde{E}_z^{*}\tilde{E}_x^*+ \tilde{E}_x\tilde{E}_z\tilde{E}_y^{*2})\\
2(\tilde{E}_y^2\tilde{E}_z^{*}\tilde{E}_y^*+ \tilde{E}_y\tilde{E}_z\tilde{E}_y^{*2})\\
2(\tilde{E}_z^2\tilde{E}_y^{*}\tilde{E}_x^*+ \tilde{E}_x\tilde{E}_y\tilde{E}_z^{*2})\\
2(\tilde{E}_z^2\tilde{E}_z^{*}\tilde{E}_x^*+ \tilde{E}_x\tilde{E}_z\tilde{E}_z^{*2}) \\
2(\tilde{E}_z^2\tilde{E}_y^{*}\tilde{E}_z^*+ \tilde{E}_z\tilde{E}_y\tilde{E}_z^{*2})\\
  \end{array}
 \right),
\]
\[
  \left(
 \begin{array}{ccccccccc}
S_{18}  \\
S_{19} \\
S_{20} \\
S_{21} \\
S_{22}\\
S_{23}\\
S_{24}\\
S_{25}\\
S_{26}\\
  \end{array}
 \right)= \left(
  \begin{array}{ccccccccc}
  4(\tilde{E}_x\tilde{E}_y\tilde{E}_z^{*}\tilde{E}_x^*+ \tilde{E}_x\tilde{E}_z\tilde{E}_x^{*}\tilde{E}_y^{*}) \\
4(\tilde{E}_x\tilde{E}_y\tilde{E}_z^{*}\tilde{E}_y^*+ \tilde{E}_z\tilde{E}_y\tilde{E}_x^{*}\tilde{E}_y^{*})\\
4(\tilde{E}_x\tilde{E}_z\tilde{E}_z^{*}\tilde{E}_y^*+ \tilde{E}_z\tilde{E}_y\tilde{E}_x^{*}\tilde{E}_z^{*})\\
i(\tilde{E}_y^2\tilde{E}_x^{*2}- \tilde{E}_x^{2}\tilde{E}_y^{*2})\\
i(\tilde{E}_z^2\tilde{E}_x^{*2}-\tilde{E}_x^2\tilde{E}_z^{*2})\\
2i(\tilde{E}_x\tilde{E}_y\tilde{E}_x^{*2}-\tilde{E}_x^2\tilde{E}_x^*\tilde{E}_y^{*})\\
2i(\tilde{E}_x\tilde{E}_z\tilde{E}_x^{*2}-\tilde{E}_x^2\tilde{E}_x^*\tilde{E}_z^{*})\\
2i(\tilde{E}_z\tilde{E}_y\tilde{E}_x^{*2}-\tilde{E}_x^2\tilde{E}_z^*\tilde{E}_y^{*})\\
i(\tilde{E}_z^2\tilde{E}_y^{*2}- \tilde{E}_y^2\tilde{E}_z^{*2})\\
\end{array}
 \right),
 \]
\begin{equation}\label{51}
 \left(
 \begin{array}{ccccccccc}
S_{27}  \\
S_{28} \\
S_{29} \\
S_{30} \\
S_{31}\\
S_{32}\\
S_{33}\\
S_{34}\\
S_{35\emph{}}\\
  \end{array}
 \right)= \left(
  \begin{array}{ccccccccc}
  2i(\tilde{E}_x\tilde{E}_y\tilde{E}_y^{*2}- \tilde{E}_y^2\tilde{E}_y^*\tilde{E}_x^{*}) \\
2i(\tilde{E}_x\tilde{E}_z\tilde{E}_y^{*2}- \tilde{E}_y^2\tilde{E}_z^*\tilde{E}_x^{*})\\
2i(\tilde{E}_y\tilde{E}_z\tilde{E}_y^{*2}- \tilde{E}_y^2\tilde{E}_z^*\tilde{E}_y^{*})\\
2i(\tilde{E}_y\tilde{E}_x\tilde{E}_z^{*2}- \tilde{E}_z^2\tilde{E}_x^*\tilde{E}_y^{*})\\
2i(\tilde{E}_x\tilde{E}_z\tilde{E}_z^{*2}-\tilde{E}_z^2\tilde{E}_x^{*}\tilde{E}_y^{*})\\
2i(\tilde{E}_y\tilde{E}_z\tilde{E}_z^{*2}-\tilde{E}_z^2\tilde{E}_z
^{*}\tilde{E}_y^{*})\\
 4i(\tilde{E}_x\tilde{E}_y\tilde{E}_z^{*}\tilde{E}_x^*- \tilde{E}_z\tilde{E}_x\tilde{E}_x^{*}\tilde{E}_y^{*})\\
 4i(\tilde{E}_z\tilde{E}_y\tilde{E}_y^{*}\tilde{E}_x^*- \tilde{E}_y\tilde{E}_x\tilde{E}_z^{*}\tilde{E}_y^{*})\\
 4i(\tilde{E}_z\tilde{E}_y\tilde{E}_z^{*}\tilde{E}_x^*- \tilde{E}_z\tilde{E}_x\tilde{E}_z^{*}\tilde{E}_y^{*})\\
 \end{array}
 \right).
 \end{equation}
From equations for incoming radiation, similar to Eq. (44) (without primes), and Eq. (51) one finds the double Stokes vector $S$ for the pure state expressed via generalized Stokes vector $s$,
\[
S_{0}=\frac{1}{\sqrt{3}}\left(-\frac{1}{2}s_1^2-\frac{1}{2}s_2^2+\frac{15}{6}s_0^2\right),
\]
\[
S_{1}=\frac{1}{\sqrt{15}}\biggl(\frac{21}{6}s_1^2-\frac{1}{2}s_2^2-\frac{3}{2}s_0^2
\]
\[
+2\sqrt{2}s_0s_1-\frac{12}{\sqrt{3}}s_1s_2+\frac{7}{\sqrt{6}}s_0s_2\biggr),
\]
\[
S_{2}=\frac{1}{\sqrt{10}}\biggl(\frac{21}{6}s_1^2-\frac{1}{2}s_2^2-\frac{3}{2}s_0^2+2\sqrt{2}s_0s_1
\]
\[
+\frac{8}{\sqrt{3}}s_1s_2-4\sqrt{\frac{2}{3}}s_0s_2\biggr),
\]
\[
S_{3}=\frac{1}{2\sqrt{6}}\left(-s_1^2+7s_2^2-4\sqrt{2}s_0s_1-3s_0^2\right),
\]
\[
S_{4}=-\frac{1}{2\sqrt{3}}s_1^2+\frac{1}{2\sqrt{3}}s_2^2+\sqrt{\frac{2}{3}}s_0s_1,
\]
\[
S_{5}=\frac{1}{\sqrt{3}}s_2\left(s_1+\sqrt{2}s_0\right),
\]
\[
S_6=\frac{1}{2}(s_3^2-s_6^2),~~~S_7=\frac{1}{2}(s_5^2-s_8^2),
\]
\[
 S_8=s_3\left(\frac{1}{\sqrt{3}}s_1+s_2+\sqrt{\frac{2}{3}}s_0\right),
\]
\[
S_{9}=s_5\left(\frac{1}{\sqrt{3}}s_1+s_2+\sqrt{\frac{2}{3}}s_0\right),~~~S_{10}=s_3s_5-s_6s_8,
\]
\[
S_{11}=\frac{1}{2}\left(s_4^2-s_7^2\right),
\]
\[
S_{12}=s_3\left(\frac{1}{\sqrt{3}}s_1-s_2+\sqrt{\frac{2}{3}}s_0\right),~S_{13}=s_3s_4+s_6s_7,
\]
\[
S_{14}=s_4\left(\frac{1}{\sqrt{3}}s_1-s_2+\sqrt{\frac{2}{3}}s_0\right),
\]
\[
S_{15}=s_5s_4-s_8s_7,~~~S_{16}=\frac{2}{\sqrt{3}}s_5\left(\frac{1}{\sqrt{2}}s_0-s_1\right),
\]
\[
S_{17}=\frac{2}{\sqrt{3}}s_4\left(\frac{1}{\sqrt{2}}s_0-s_1\right),
\]
\[
S_{18}=2s_4\left(\frac{1}{\sqrt{3}}s_1+s_2+\sqrt{\frac{2}{3}}s_0\right),
\]
\[
S_{19}=2s_5\left(\frac{1}{\sqrt{3}}s_1-s_2+\sqrt{\frac{2}{3}}s_0\right),
\]
\[
S_{20}=\frac{4}{\sqrt{3}}s_3\left(\frac{1}{\sqrt{2}}s_0-s_1\right),
\]
\[S_{21}=-s_3s_6,~~~S_{22}=-s_5s_8,
\]
\[
S_{23}=-s_6\left(\frac{1}{\sqrt{3}}s_1+s_2+\sqrt{\frac{2}{3}}s_0\right),
\]
\[
S_{24}=-s_8\left(\frac{1}{\sqrt{3}}s_1+s_2+\sqrt{\frac{2}{3}}s_0\right),
\]
\[
S_{25}=-s_3s_8-s_5s_6,~~~S_{26}=-s_4s_7,
\]
\[
S_{27}=s_6\left(\frac{1}{\sqrt{3}}s_1-s_2+\sqrt{\frac{2}{3}}s_0\right),
\]
\[
S_{28}=-s_4s_6-s_3s_7,
\]
\[
S_{29}=-s_7\left(\frac{1}{\sqrt{3}}s_1-s_2+\sqrt{\frac{2}{3}}s_0\right),
\]
\[
S_{30}=s_4s_8+s_5s_7,
\]
\[
S_{31}=\frac{2}{\sqrt{3}}s_8\left(\frac{1}{\sqrt{2}}s_0-s_1\right),
\]
\[
S_{32}=\frac{2}{\sqrt{3}}s_7\left(\frac{1}{\sqrt{2}}s_0-s_1\right),
\]
\[
S_{33}=2s_7\left(\frac{1}{\sqrt{3}}s_1+s_2+\sqrt{\frac{2}{3}}s_0\right),
\]
\begin{equation}\label{52}
S_{34}=-s_4s_6-s_3s_7,~~S_{35}=2\sqrt{\frac{2}{3}} s_6(\sqrt{2}s_1-s_0).
\end{equation}

\section{Appendix C: Rotational transformations}

After rotational transformations, the 3-component vector $\Omega$ in Eq. (2) becomes (for simplicity we omit primes) $\Omega_R=R\Omega$, where $3\times 3$ rotation $R$-matrix is orthogonal, $RR^T=R^TR=I$ (see, for example \cite{Kruglov5}). From Eq. (4), we find the new rotated Stokes vector: $s_{AR}=\langle \Omega^\dag R^\dag\lambda_AR\Omega \rangle$. Note that $R^\dag =R^T$ because the matrix $R$ is real ($R^*=R$). We can introduce ``rotational" matrix $\lambda_{AR}=R^\dag\lambda_AR$ so that $s_{AR}=\langle \Omega^\dag\lambda_{AR}\Omega \rangle$. Thus, we can interpret the rotation as changing the matrices $\lambda_A$ to $\lambda_{AR}$ but leaving the vector $\Omega$ to be unchanged.

We introduce new $9$-component vector $\Psi=\Omega\otimes\Omega$. Under the rotation $\Psi_R=R\Omega\otimes R\Omega=(R\otimes R)(\Omega\otimes \Omega)$. Thus, the rotation in $9$-dimensional space is given by the orthogonal matrix $R\otimes R$ so that
$(R\otimes R)^T=R^T\otimes R^T$. Then, after the rotation, double Stokes vector becomes $ S_{NR}=\Psi^\dag (R\otimes R)^\dag\eta_N(R\otimes R)\Psi=\Psi^\dag \eta_{NR}\Psi$, where the ``rotational" matrix is $\eta_{NR}=(R\otimes R)^\dag \eta_N (R\otimes R)$. Here the $9\times 9$ matrix $\eta_N$ was introduced in Eq. (11) instead of $6\times 6$ matrix $\eta_N$ defined by Eq. (50).
Then, after the rotation, the $27\times 27$ matrix $X$ becomes
\[
X_R=\frac{1}{2} {\cal M}_{tN}\lambda_{tR}\otimes \eta_{NR}^T
\]
\[
=\frac{1}{2} {\cal M}_{tN}R^T\lambda_{t}R\otimes
((R\otimes R)^T\eta_{N}^T(R\otimes R))
\]
\begin{equation}\label{53}
=(R\otimes R\otimes R)^TX(R\otimes R\otimes R).
\end{equation}
As the matrix $K\equiv R\otimes R\otimes R$ is orthogonal, $(R\otimes R\otimes R)(R\otimes R\otimes R)^T=1$, eigenvalues of the $27\times27$ matrix $X$ are rotational invariant because the characteristic equation is
\[
\mbox{det}(X_R-\kappa I)=\mbox{det}(K^T(X-\kappa I)K)
\]
\begin{equation}\label{54}
=\mbox{det}(KK^T(X-\kappa I))=\mbox{det}(X-\kappa I)=0.
\end{equation}
Here we make use of the equality $KK^T=1$. Eigenvalues of $X$-matrix characterize the sources of the outgoing radiation and, therefore, the rotation invariance of eigenvalues is important.

\section{Funding Information}

This work was supported by Natural Sciences and Engineering Research Council of Canada (RGPIN-2017-06923, DGDND-2017-00099).


\begin{thebibliography}{99}

\bibitem{Kruglov} M. Samim, S. Krouglov and V. Barzda, J. Opt. Soc. Am. B \textbf{32}, 451 (2015);
 Phys. Rev. A \textbf{93}, 013847 (2016); ibid \textbf{93}, 033839 (2016).

\bibitem{Verbiest} T. Verbiest, K. Clays, and V. Rodriguez \textit{Second-Order Nonlinear Optical CharacterizationTechniques}
(CRCPress, 2009).

\bibitem{Simpson} G. J. Simpson, J. Garth, \textit{Nonlinear Optical Polarization Analysis in Chemistry and Biology} (Cambridge University Press, 2017).

\bibitem{Clays} K. Clays and A. Persoons, Phys. Rev. Lett. \textbf{66}, 2980 (1991).

\bibitem{Maker} P. D. Maker, Phys. Rev. A \textbf{1},923 (1970).

\bibitem{Shi} Y. Shi, W. M. McClain, and R. A. Harris, Phys. Rev. A \textbf{49}, 1999 (1994).

\bibitem{Zhuang} X. Zhuang, P. B. Miranda, D. Kim, and Y. R. Shen, Phys. Rev. B \textbf{5}, 12632 (1999).

\bibitem{Tuer} A. E. Tuer, M. K. Akens, S. Krouglov, D. Sandkuijl, B. C. Wilson, C. M. Whyne, and V. Barzda, Biophys. J. \textbf{103}, 2093 (2012).

\bibitem{Loew} P. J. Campagnola and L. M. Loew, Natl. Biotechnol. \textbf{21}, 1356 (2003).

\bibitem{Plotnikov} S. Plotnikov, A. Millard, P. Campagnola, and W. Mohler, Biophys. J. \textbf{90}, 693 (2006).

\bibitem{Nucciotti} V. Nucciotti, C. Stringari, L. Sacconi, F. Vanzi, L. Fusi, M. Linari, G. Piazzesi, V. Lombardi, and F. S. Pavone, Proc. Natl. Acad, Sci. USA \textbf{107}, 7763 (2010).

\bibitem{Vanzi} F. Vanzi, L. Sacconi, R. Cicchi, and F. S. Pavone, J. Biomed. Optics \textbf{17} (2012).

\bibitem{Golaraei} A. Golaraei, R. Cisek, S. Krouglov, R. Navab, C. Niu, S. Sakashita, K. Yasufuku, M. S. Tsao, V. C. Wilson, andV. Barzda, Biomed. Optics. Express \textbf{5}, 3562 (2014).

\bibitem{Kontenis} L. Kontenis, M. Samim, A. Karunendiran, S. Krouglov, B. Stewart, and V. Barzda, Biomed. Opt. Express, \textbf{7}, 559 (2016).

\bibitem{Kontenis1} L. Kontenis, M. Samim,  S. Krouglov, and V. Barzda, Opt. Express, \textbf{25}, 13174 (2017).

\bibitem{Kliger} D. S. Kliger, J. W. Lewis, and C. E. Randall, \textit{Polarized Light in Optics and Spectroskopy} (Academic Press, 1990).

\bibitem{Shurcliff} W. A. Shurcliff, \textit{Polarized light: Production and Use} (Harvard University Press, Cambrige, MA, 1962).

\bibitem{Lien} C.-H. Lien, K. Tilbury, S.-J. Chen, and P. Campagnola, Biomed. Optics. Express \textbf{4}, 1991 (2013).

\bibitem{Mazumder} N. Mazumder, C.-W. Hu, J. Qiu, M. R. Foreman, C. M. Romero, P. T\"{o}r\"{o}k, and F.-J. Kao, Methods, \textbf{66}, 237 (2014).

\bibitem{Mazumder1} N. Mazumder, J. Qiu, M. R. Foreman, C. M. Romero, C.-W. Hu, H.-R. Tsai,  P. T\"{o}r\"{o}k, and F.-J. Kao, Optics. Express \textbf{20}, 14090 (2012).

\bibitem{Mazumder2} N. Mazumder, J. Qiu, M. R. Foreman, C. M. Romero,  P. T\"{o}r\"{o}k, and F.-J. Kao, Biomed. Optics. Express \textbf{4}, 538 (2013).

\bibitem{Samson} J. C. Samson, Geophys. J. R. Ast. Soc. \textbf{34}, 403 (1973).
%Descriptions of the polarization states of vector processes: applications to ULF magnetic fields -19

\bibitem{Barakat}  R. Barakat, Opt. Comm. \textbf{23}, 147 (1977).
%Degree of polarization and the principal idempotents of the coherency matrix –50

\bibitem{Samson1} J. C. Samson and J. V. Olson, Geophys. J. R. Ast. Soc. \textbf{61}, 115 {1980}.
%Some comments on the descriptions of the polarization states of waves -29

\bibitem{Kaivola1} T. Set\"{a}l\"{a}, M. Kaivola, and A. T. Friberg, Phys. Rev. Lett. \textbf{88}, 123902 (2002).
%Degree of polarization in near fields of thermal sources: effects of surface waves

\bibitem{Kaivola}  T. Set\"{a}l\"{a}, A. Shevchenko, M. Kaivola, and A. T. Friberg, Phys. Rev. E \textbf{66}, 016615 (2002).
%Degree of polarization for optical near fields

\bibitem{Brosseau_0}  C. Brosseau and A. Dogariu, Progress in Optics, \textbf{49}, 315 (2006).
%Symmetry properties and polarization descriptors for an arbitrary electromagnetic wavefield

\bibitem{Brosseau} C. Brosseau, \textit{Fundamentals of polarized light: a statistical optics approach} (Wiey, New York, 1998).

\bibitem{Dennis} M. R. Dennis, J. Opt. A: Pure Appl. Opt. \textbf{6}, 26 (2004).

\bibitem{Gil} J. J. Gil, Phys. Rev. A \textbf{90}, 043858 (2014);
J. Eur. Opt. Soc.-Rapid  \textbf{10}, 15054 (2015).

 \bibitem{Boyd} R. Boyd, \textit{Nonlinear Optics}, 3rd ed. (Academic Press, Boston, 2008).

 \bibitem{Kruglov4} S. I. Kruglov, \textit{Symmetry and electromagnetic interaction of fields with multi-spin.
A Volume in Contemporary Fundamental Physics} (New York: Nova Science Publishers, Huntington, 2001).

\bibitem{Kruglov1} M. Samim, S. Krouglov, D. F. James and V. Barzda, J. Opt. Soc. Am. B \textbf{33}, 2617 (2016).

\bibitem{Gil2} J. J. Gil, J. Opt. Soc. Am. A \textbf{17}, 328 (2000); Eur. Phys. J. Appl. Phys. \textbf{40}, 1 (2007).

\bibitem{Gil1} J. J. Gil and E. Bernabeu, Opt. Acta \textbf{33}, 185 (1986).

\bibitem{Roy} F. Le Roy-Brehonnet and Le Jeune, Prog. Quant. Electr. \textbf{21}, 109 (1997).

\bibitem{James} D. F. V. James, P. G. Kwiat, W. J. Munro, and A. G. White, Phys. Rev. A \textbf{64}, 052312 (2001).

\bibitem{James1} M. S. Kaznady and D. F. V. James, Phys. Rev. A \textbf{79}, 022109 (2009).

\bibitem{Cholesky} A.-L. Cholesky, bulletin de la soci\'{e}t\'{e} des amis de la biblioth\'{e}qu\'{e} de I\'{E}cole polytechnique (SABIX) \textbf{39}, (1910).

\bibitem{Sheppard} C. J. R. Sheppard, Phys. Rev. A \textbf{90}, 023809 (2014).

\bibitem{Kruglov5} S. I. Kruglov and V. Barzda, Annales Fond. Broglie \textbf{42}, 235 (2017).

\bibitem{Carozzi} T. Carozzi, R. Karlsson, and J. Bergman, Phys. Rev. E \textbf{61}, 2024 (2000).

\end{thebibliography}
\end{document}